\documentclass[journal]{IEEEtran}
\usepackage{subfig}
\usepackage{amsmath}
\usepackage{graphicx}
\usepackage{float}
\usepackage{nomencl}
\usepackage{blindtext}
\usepackage{multicol}
\usepackage{caption}
\usepackage{flushend}
\usepackage[numbers,sort&compress]{natbib}
 \usepackage{multirow}
\usepackage{diagbox}
\usepackage{multicol}

\ifCLASSINFOpdf
 
\else
 
\fi

\begin{document}

\title{Smart Inverters' Functionalities and their Impacts on Distribution Feeders at High Photovoltaic Penetration: An Overview}

\author{Temitayo~O. Olowu,~\IEEEmembership{Student Member,~IEEE,}
        Arif~Sarwat,~\IEEEmembership{Senior Member, ~IEEE,}
\thanks{T. O Olowu and A. I. Sarwat are with the Department
of Electrical and Computer Engineering, Florida International University, Miami, United States. }}

\maketitle

\begin{abstract}

The increase in grid-connected photovoltaic (PV) systems has been consistent over the years. Several studies have reported the impacts of this increase on the distribution networks such as reverse power flow, voltage fluctuations, possible increase in system losses amongst others. With this rise in PV penetration level, it becomes necessary for grid tied smart inverters (SI) to be allowed to participate in feeder voltage regulation. Till date, no comprehensive technical review and analysis has been done on the impact and interaction of existing and proposed SI settings on the existing legacy voltage control devices. This paper presents a novel technical review and analysis of some voltage regulation settings of SIs and their  impacts on distribution feeders including their interaction with legacy devices such as capacitors, on/off Load Tap Changers, their impact on the feeder losses, harmonic effects and economic impacts (expressed as the device cost factor) and circuit impact index (CII). 
The analysis of these functionalities is done using the standard IEEE $8500$ distribution feeder, integrated with six PVs (based on real PV parameters and data), and strategically located with actual irradiance and temperature profile from a 1.4MW PV plant located at FIU. These PV locations were carefully selected to allow for detailed impacts studies. The results show how the various SI functions impact the reactive power injection and switching of the capacitor banks, the voltage regulator switching, the losses in the feeder, the harmonic and the CII.  The Volt-Watt with rise/fall rate-of-change limiting  setting showed the least impact in terms of the CII, voltage regulator tapping and capacitor switching but with high amount of losses compared to other SI functions.
\end{abstract}


\IEEEpeerreviewmaketitle

\section{Introduction}

\IEEEPARstart{T}{he} total power generation from photovoltaic (PV) systems globally is expected to reach over $1 TW$ by 2022 according to \cite{europe2018global}. With the continued increase in level of PV penetration, its consequent challenges have to be addressed \cite{TayoISGT,ariyo2016web,8479058,olowu2018future}. SI technologies, obviously will play a huge role in addressing some of these issues as result of the increase in penetration level of PV systems on the smart grid \cite{8440431,8107405,sarwat2017trends,TayoIAS}.
The use of SI functions have gained a lot attention recently. This is obviously due to the increasing level of PV penetration on the grid at various voltage levels. In its simplest form, a SI is defined as a power electronic device that converts a DC power to an AC \cite{Jafarian2017,8536872} with monitoring, communication and control capabilities. Modern SIs are intelligent microprocessor-based which allows for Ethernet/LAN/WAN connectivity \cite{Carvallo2011}. 

According to \cite{Xue2015}, some advanced desirable capabilities of SIs include self-optimization, self-configuration, self-awareness, plug-and-play, adaptability, autonomy, cooperativeness. SIs (also sometimes referred to as advanced inverters) enables comprehensive monitoring of the grid status, remote communication and control from command and control center, improvement of grid stability through autonomous decisions, power quality control and provision of other support services \cite{NationalRenewableEnergyLaboratory2014}. 

The use of smart inverters is fast becoming an attractive option for grid voltage optimization. This is due to their ability for fast  and flexible control of their real (curtailment and ramp control) and reactive power (injection and absorption) \cite{Schauder2014}. Also, selection of the optimal settings of these smart inverters could help reduce the number of switching of voltage regulators (VR), on/off load tap changers, and capacitors banks. As a consequence, this will help extend the life span of these legacy devices and reduce the cost of maintenance. An NREL report \cite{Ding2016}, demonstrated how the use of smart inverters can help utility companies better implement a conservative voltage reduction (CVR) energy savings. The implementation of CVR can be integrated as part of the voltage optimization algorithms. The IEEE 1547a-2014 \cite{6818982} is the first standard that recommended that smart inverter based DER should actively participate in voltage control and regulation by changing their reactive or real power \cite{Huque2015}. Conventionally, voltage regulators, switching capacitors (or shunt capacitors), on/off load tap changers, and reactors are used for voltage regulation and control in traditional distribution grids. Table \ref{fig:table 1} shows some important smart inverter functionalities as recommended by the  newly amended IEEE 1547-2018 \cite{IEEE15472018,olowu2018future}. 

At high penetration of SI-based DERs, voltage regulation, voltage stability, reverse Power flow, complexity in protection coordination, harmonics, frequency instability, feeder losses, amongst others, have become a huge challenge \cite{olowu2018future}.  In order to address these issues, the IEEE 1547-2018 \cite{8332112} standard requires (based on the level of penetration described as categories A and B) the SIs to have voltage and frequency ride-through capabilities. This will prevent a cascading of DER disconnection from the grid during abnormal occurrences. Several papers \cite{7320704,8447750,kim2016new} have presented studies on how the delay in the voltage and frequency control loop of the SI voltage control functions could cause a voltage instability in the system. Therefore it is required that SI dynamic response should be fast enough to act during abnormal conditions to address these issues. The required time of operation/response during these abnormal scenarios (categorized as category I, II and III) is as stipulated in the IEEE 1547-2018 standard. 

Since the voltage control and optimization functions of SIs impact the feeder dynamics differently, technical review of these SI functions becomes imperative with increasing demand of its use by power utility companies.

The key contributions of this paper includes (a) A novel detailed technical review of the impacts of voltage control SI functions (volt-Watt, volt-Watt with rise/fall rate-of-change limit, volt-VAR, volt-VAR with adaptive set-point, volt-VAR with hysteresis, volt-VAR with low pass filter rate-of-change limit, maximum generation limit at 80 \%,  fixed power factor (FPF) at $0.8$, dynamic reactive current injection/absorption) on operations of voltage control devices (b) Economic impact analysis using circuit impact index and (c) the harmonic impact of these SI functions on the feeder under study. 
The rest of this paper is organized as follows: Section II presents a brief review of the related works to this paper;
Section III  presents a brief description of some smart inverter functions considered,
Section IV describes the method used to carry out this simulation and analysis; Section V presents the simulation results and analysis of the results while Section VI concludes the paper.

\begin{table}[h!]
\caption{Smart Inverter Functionalities.}
\centering
\footnotesize
\label{fig:table 1}
\begin{tabular}{|m{2.5cm}|m{2cm}|m{3cm}|}
\hline
 \centering{\textbf{ Functionalities}}& \centering{\textbf{Sub-Functionalities}} & \textbf{Specific Settings}  \\
 \hline
\centering{Voltage Ride Through (VRT)} & \centering{Low/High VRT}& Voltage, Duration (time)\\
\hline
\centering{Frequency Ride-Through (FRT)}&	\centering{Low/High FRT}	&Frequency, Duration (time)\\
\hline
\centering {Dynamic Volt-VAR/Watt Control ramping}&	\centering{Volt-VAR, Volt-Watt} &Volt-VAR/Watt Curves, Ramp rates\\
\hline
\multicolumn{2}{|c|}{Power Factor setting/control}&Values\\
\hline
\centering{Soft start}	& \multicolumn{2}{c|}{Ramp rate, Time delay}\\

\hline
\centering{Limit Real and Reactive Power}&\multicolumn{2}{c|}{Enable/Disable}\\
\hline
\centering{Frequency-Watt}&&	Frequency-Watt Curve \\
\hline
\multicolumn{3}{|c|}{Dynamic Current Support}\\
\hline
\centering{Output Scheduling} & \multicolumn{2}{m{5.5cm}|}{Time of start, Time to end, Real and Reactive power value, operational schedule}

\\
\hline

\multicolumn{3}{|c|}{Frequency Deviation Support}\\
\hline
\multicolumn{3}{|c|}{Control of Reactive Power Dynamically}\\
\hline
\multicolumn{3}{|c|}{Dynamic Load Control}\\
\hline

\multicolumn{3}{|c|}{Dynamic Harmonic Control}\\
%\centering{Dynamic Harmonic Control}&&\\
\hline
\end{tabular}
\end{table}

\section{Related Works} 

Several studies have been carried out to analyze the various impacts of SI functions on the grid. Some of these studies were carried out using specific utility feeders. A report by \cite{Nelson2016} presented the validation of SI function on the feeder voltage profile using two feeders from the Hawaiian distribution network. This study carried out a laboratory test on five SIs (with grid support functions) using power hardware-in-the-loop. The study presented the impacts of the FPF and Volt-Watt curve (VWC) on voltage profiles of the feeders. The study showed that the FPF had some significant impacts in reducing the high voltages of the network compared  to VWC. The study also  showed that at higher levels PV integration, the effectiveness of this voltage reduction using the FPF becomes more effective compared to the VWC. A study done by \cite{Pecenak2018} showed that the variability in the feeder voltage profile and the tapping by voltage regulators can be improved with the use of SIs and also, a volt-VAR control setting of SIs without the dead-band  improves the effectiveness of the voltage control. This paper only presented the potential benefits of the volt-VAR control. Authors of \cite{7749843} described some methods by which distribution planning engineers can carefully select a SI settings that can improve the performance of distribution feeders.  This paper simulated the possible benefits of some SI functions (volt-VAR, volt-Watt and FPF) such as improved hosting capacity and reduction in the system losses. Reference \cite{Dubey2017} used the EPRI $34.5 kV$ test circuit to analyze the impact of the volt-VAR and FPF settings for voltage control with and without the use of the existing legacy voltage control devices. The paper showed that there is an increase in the reactive power demanded by the system when the volt-VAR setting was used.

Also, Authors of \cite{8598813} presented a Volt-VAR optimization technique that controls the voltage control operations of feeder's legacy devices and coordinates their operation with SIs. The voltage control is done in order to achieve conservative voltage reduction which reduces the total load demand by simply controlling the nodal voltages of the feeder. A two-level Volt-VAR optimization was proposed. The first level optimizes the legacy devices and SI control operation by using a three phase load flow with linear approximation, while the second level involves, adjusting the the control of the SI in order to achieve a feasible nonlinear OPF solution. Authors of \cite{8759872} proposed the use of a model predictive control which is based on support vector regression machine learning algorithm for Volt-VAR optimization. The MPC is used to control the capacitor banks and the tap changers while minimizing the system's power loss. The proposed technique has the benefit of not requiring any simulation results from the feeder. The SVR is trained using data collected from an advanced metering infrastructure (AMI). Reference \cite{Pamshetti} proposed a local and centralized VVO algorithm that minimizes the total operating cost and node voltage deviation while implementing CVR. The optimization problem was solved using the fuzzy logic algorithm and the $\varepsilon$-constraint method. The proposed VVO was validated on an IEEE 33-bus system which showed the effectiveness of the proposed approach.

However, these cited works amongst others, did not consider many other SI functionalities as well as a detailed impact studies.
Most of the studies carried out on SI functions, only demonstrates the effectiveness and dynamic operations of these functions, little has been reported on how these functions affect the operation of the legacy devices in the network, the feeder losses, the potential economic impacts on these devices as well as their power quality impact on distribution feeders which very important in selecting the appropriate functions depending in the feeder characteristics and the level of PV penetration.

\section{Brief Description of some voltage control smart inverter functions}
There are numerous functions capable of being executed with the use of smart inverters. Some of these setting have a direct impact on the feeder performance as well as the operations of the voltage control devices on the network according to \cite{EPRI2016,Smith2013}. A brief description of some of these functions is presented below.
\subsection{Constant Power Factor (PF) Setting}
\begin{figure}[h!]
    \centering
\includegraphics[scale=0.16]{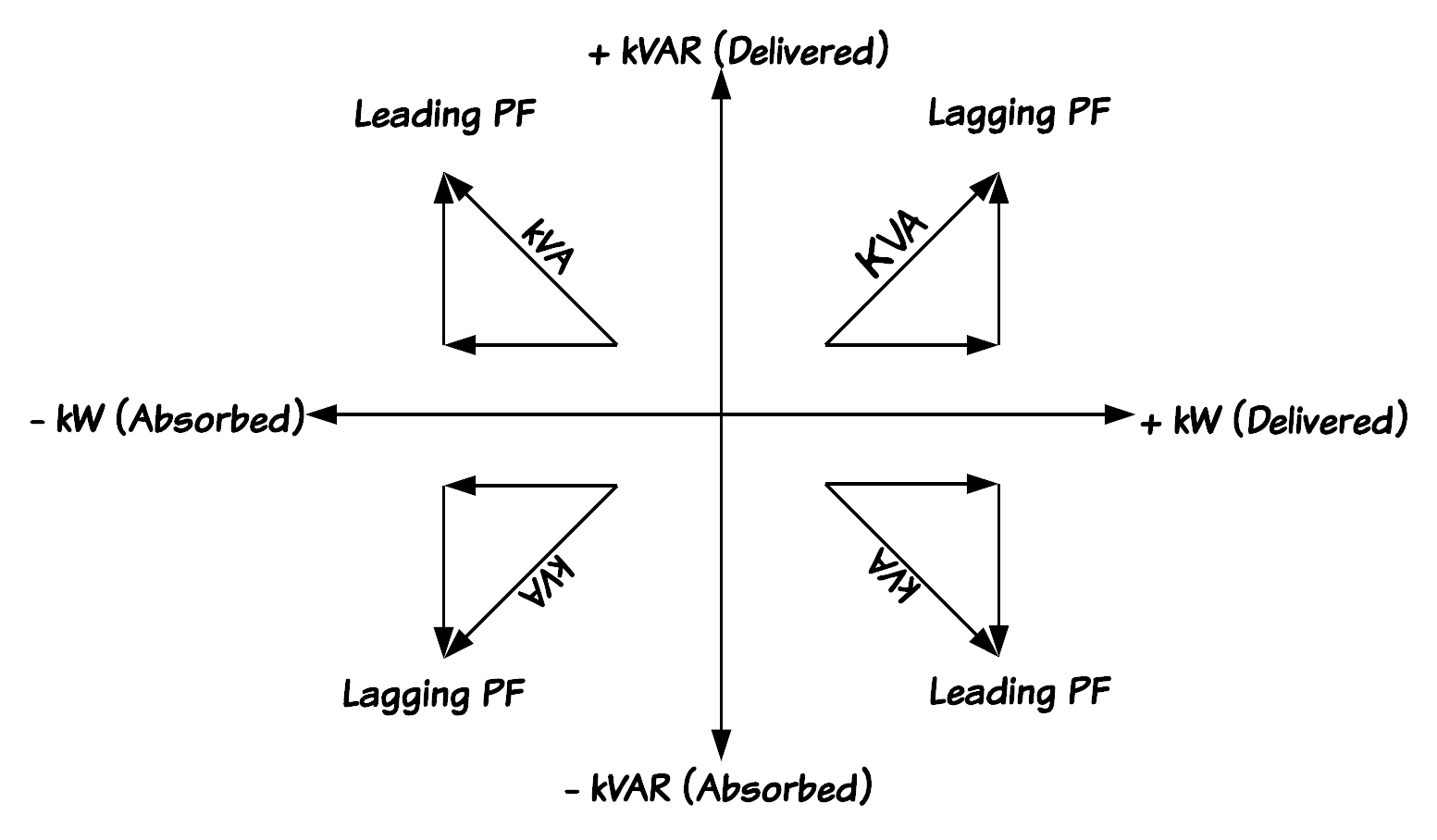}
    \caption{Power factor quadrants}.
    \label{fig:PFQ}
\end{figure}

The power factor setting is the basic and primarily the most widely used smart inverter settings. It could vary from -1.00 to +1.00. Usually, a zero PF setting (100 \% reactive power injection/absorption) is not allowed \cite{IEEE15472018}. A PF value of 1 allows the PV to export all its active power generation (kW) to the grid. This is typical of most PV producers’ smart inverters, in order to maximize their revenue from PV generation. Figure \ref{fig:PFQ} shows the four possible operating points for a fixed power factor settings on a smart inverter.
\subsection{Volt-VAR Control}
\begin{figure}[h!] 
  \begin{center}
 \subfloat[]{\label{fig:VVC}\includegraphics[scale=0.2]{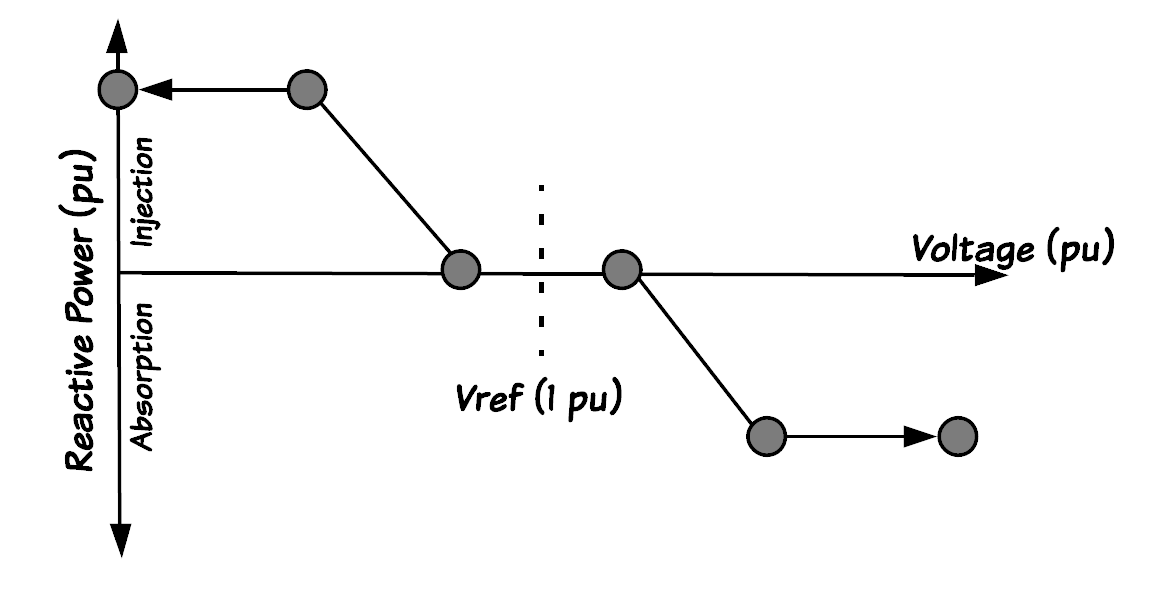}}\hspace{1cm}
  \subfloat[]{\label{VVC-Hys}\includegraphics[scale=0.2]{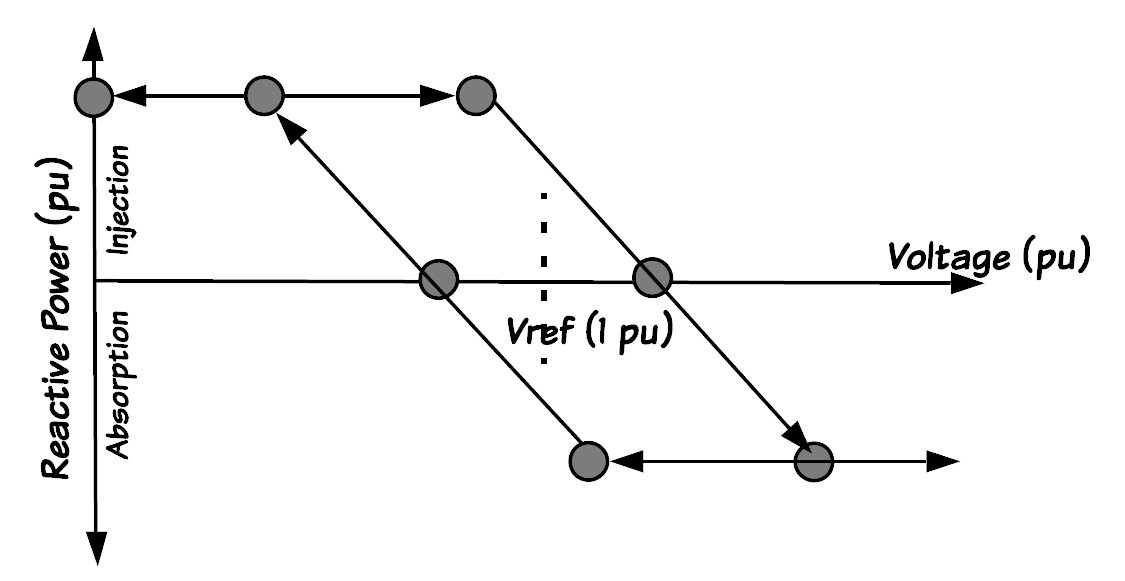}}
    \caption{\label{fig:VVC-Hys} Typical Volt-VAR curves operating points  (a)Without an hysteresis band (b) With an hysteresis band}
  \end{center}
\end{figure}

The Volt-VAR functionality, is the most discussed setting of the smart inverter. This is due to the ease at which a voltage regulation can be achieved with reactive power control.  The reactive injection or absorption by the inverter is determined by the terminal voltage at the point of interconnection of the inverter. A typical operating point for the Volt-VAR control is shown in Fig. \ref{fig:VVC}. A hysteresis band can be added to the Volt-VAR curve as shown in Fig. \ref{VVC-Hys}. The inclusion of the hysteresis band allows for a constant reactive power output when the voltage rise or fall is not high enough to encounter the points set  by the Volt-VAR curves.

\subsection{Volt-Watt Control}
\begin{figure}[h!] 
  \begin{center}
 \subfloat[]{\label{fig:VWC1}\includegraphics[scale=0.2]{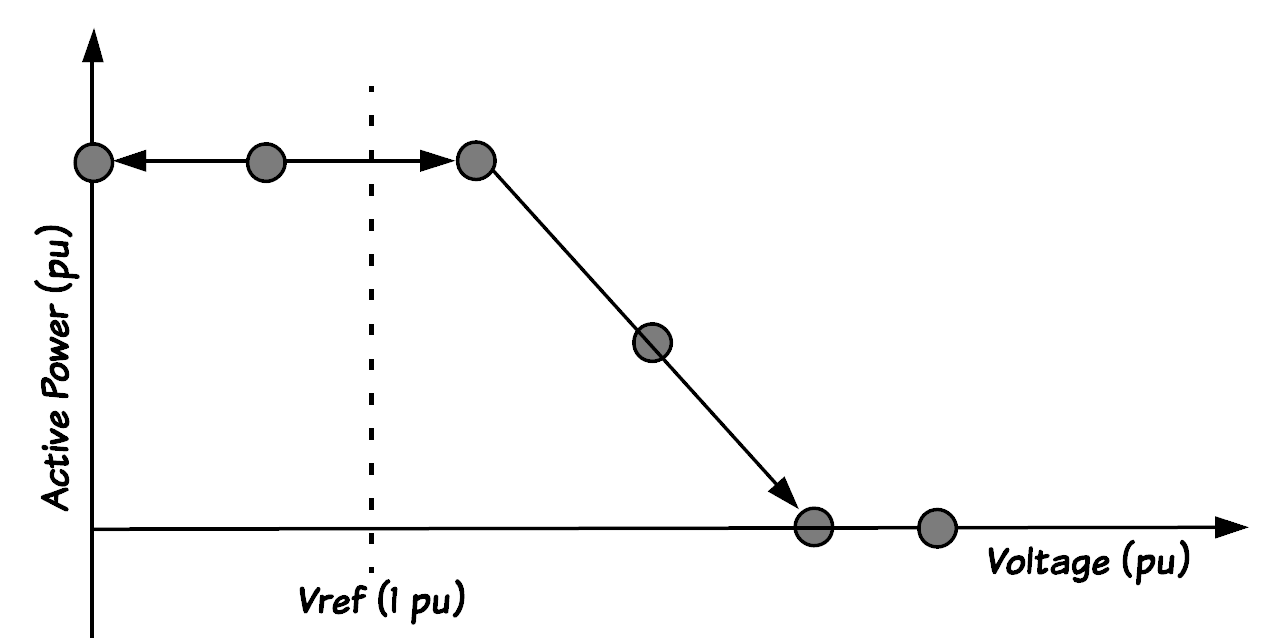}}\hspace{1cm}
  \subfloat[]{\label{fig:VWC2}\includegraphics[scale=0.2]{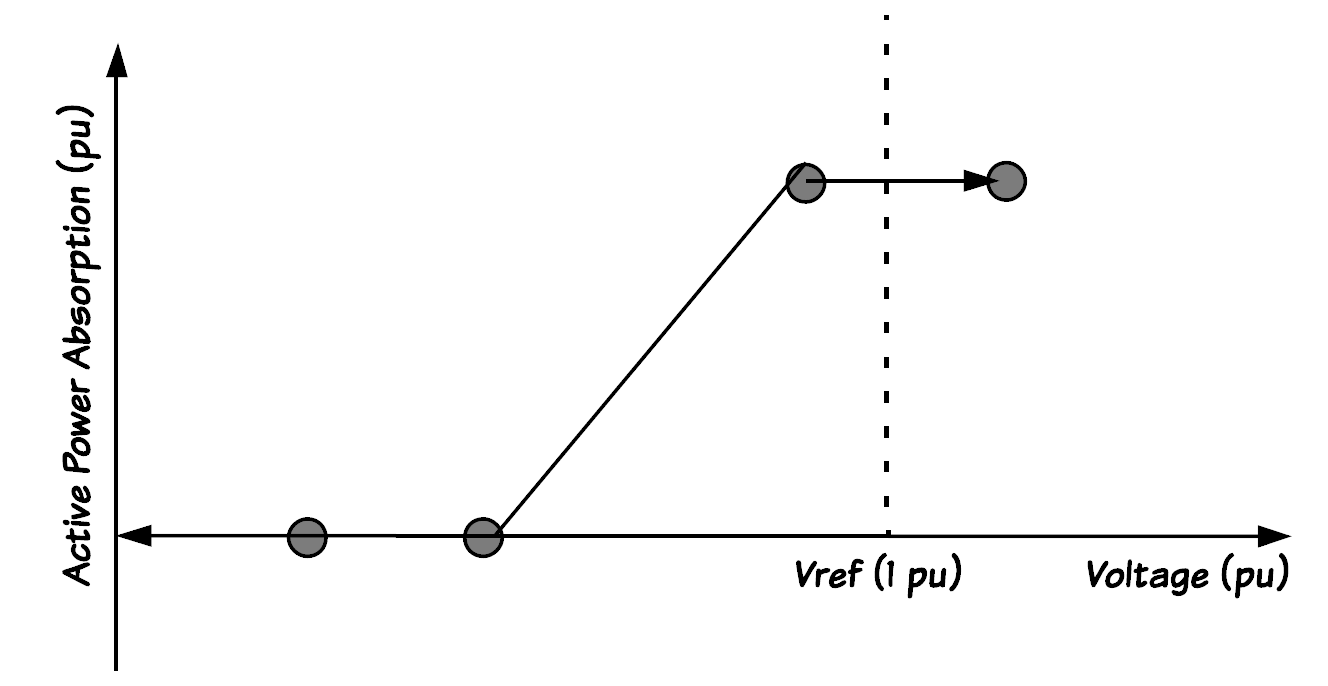}}
    \caption{\label{fig:VWC} Volt-Watt curves (a) Active power generation in response to the terminal voltage (b) Active power absorption (by possible energy storages) in response to the terminal voltage}
  \end{center}
\end{figure}

The smart inverter Volt-Watt setting controls the active power generation by the PV system in response to the terminal voltage at the point of interconnection \ref{fig:VWC}. The rate of active power reduction based on the voltage at the point if interconnection is determined by the slope of the curve as shown in Fig. \ref{fig:VWC1}. Also, the rate of absorption of active power by storage element can be defined using the volt-Watt curve as shown in Fig. \ref{fig:VWC2}.

\subsection{Frequency-Watt Control}
\begin{figure}[ht] 
  \begin{center}
 \subfloat[]{\label{fig:FW}\includegraphics[scale=0.18]{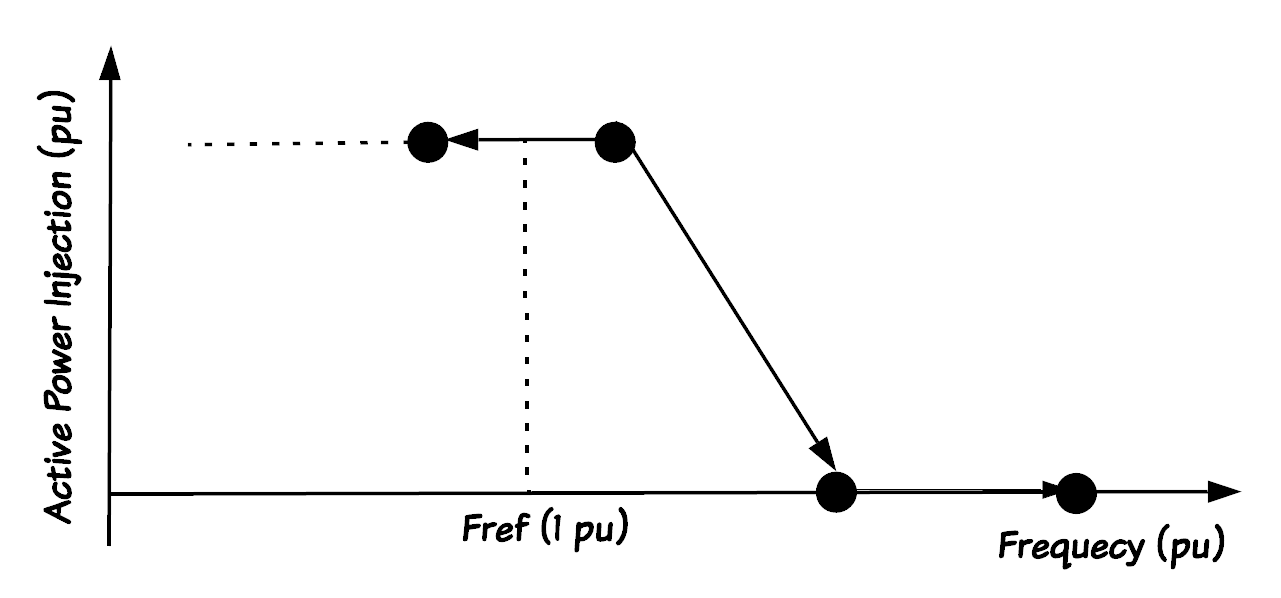}}\hspace{1cm}
  \subfloat[]{\label{fig:FW-Hsy}\includegraphics[scale=0.18]{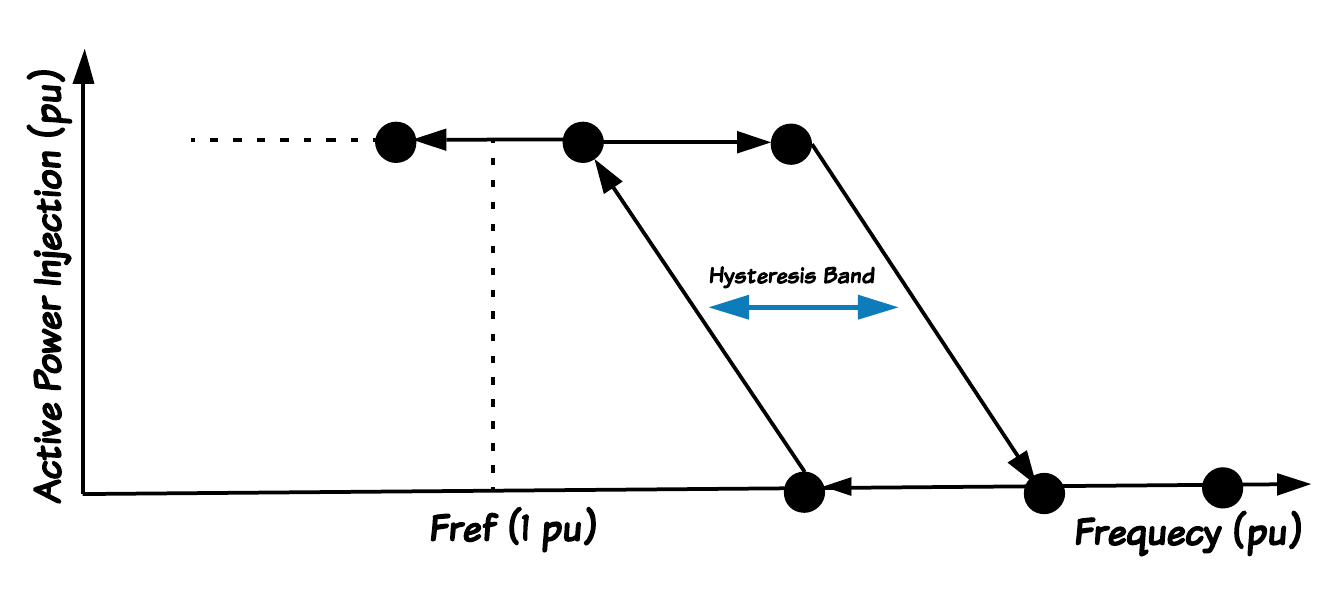}}\hspace{1cm}
   \subfloat[]{\label{fig:FW-Hsy_2}\includegraphics[scale=0.18]{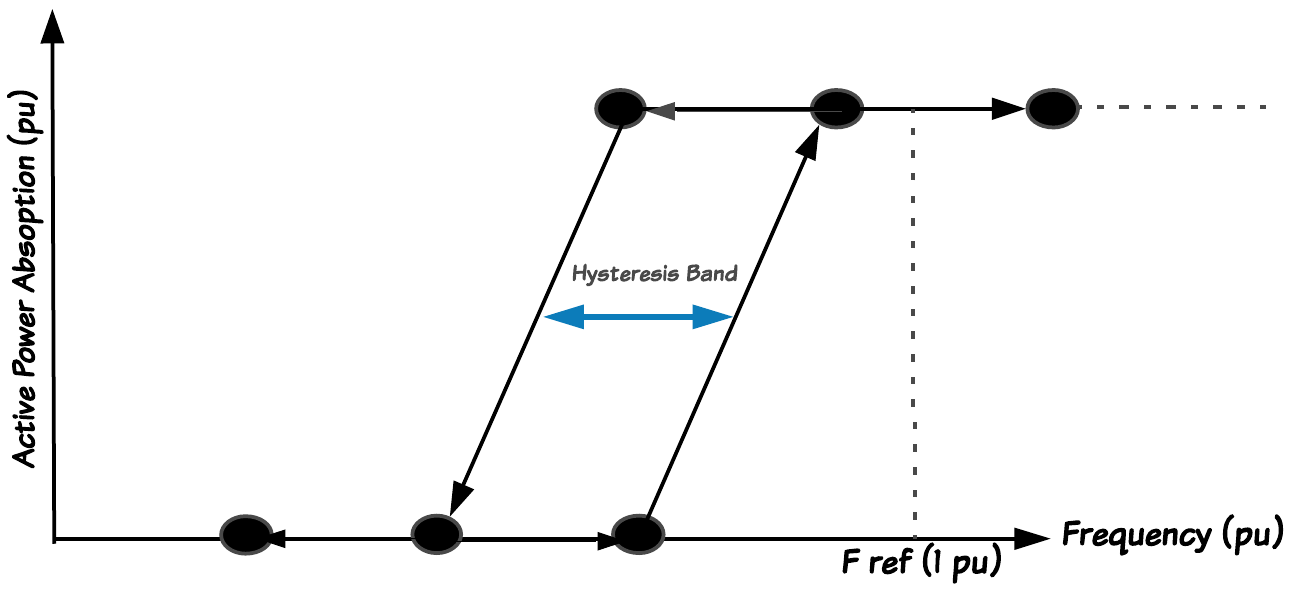}}
    \caption{\label{fig:FW3} Frequency-Watt curves (a) Active power generation in response to the terminal voltage (b) Active power absorption (by possible energy storages) in response to the terminal voltage}
  \end{center}
\end{figure}

The frequency-watt control function allows the inverter to control the active power generation in response to the system frequency \cite{hoke2017frequency,7468132}. The operation of the smart inverter using this control function is similar to that of the droop control done by the governor of synchronous generators. The typical operating points for this function is shown in Fig. \ref{fig:FW}. Figures \ref{fig:FW-Hsy} and \ref{fig:FW-Hsy_2} show a frequency watt control curve with an hysteresis band and a two-way power flow (injection and absorption). This bidirectional power flow functionality allows the frequency to be reduced by absorbing active power (using energy storage) and be increased by active power injection.

\subsection{Maximum Generation Limit Function}
This function often referred to as active power curtailment is used to create a set point for maximum amount of active power that can be supplied by the PV system through the  smart inverter. It is usually set as a percentage of the maximum power available from the PV at every point in time.

\subsection{Dynamic Reactive Current Injection}
\begin{figure}[h!]
    \centering
\includegraphics[scale=0.15 ]{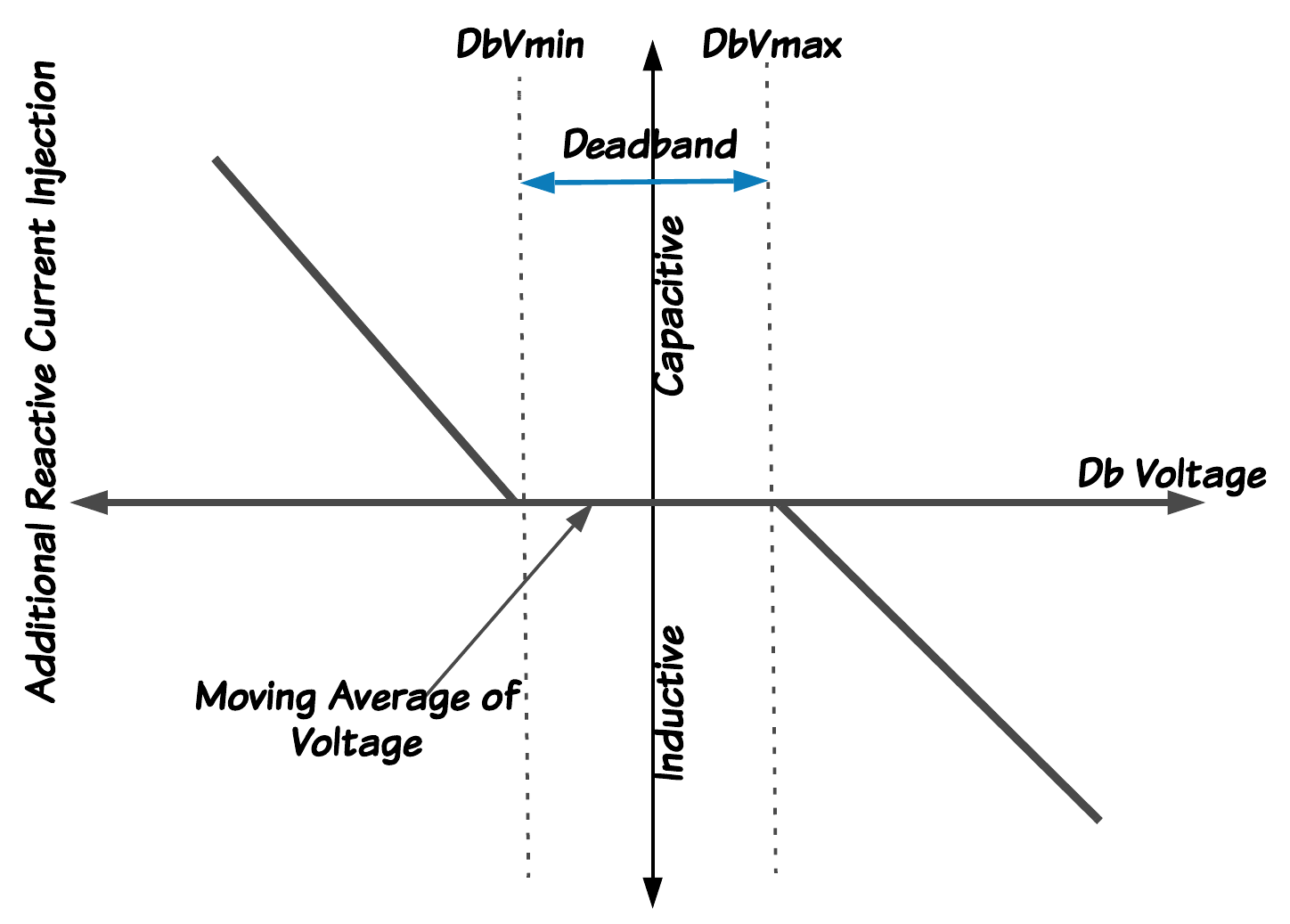}
    \caption{A typical curve for reactive current injection }.
    \label{fig:DCI}
\end{figure}

This function of the smart inverter is similar to the Volt-VAR control. But rather than changing the reactive power injection, the smart inverter generates some reactive current in response to the changes in the voltage. A typical curve depicting the dynamics of this functionality is shown in \ref{fig:DCI}.

\subsection{Watt-Power Factor Function}
\begin{figure}[h!]
    \centering

\includegraphics[scale=0.23]{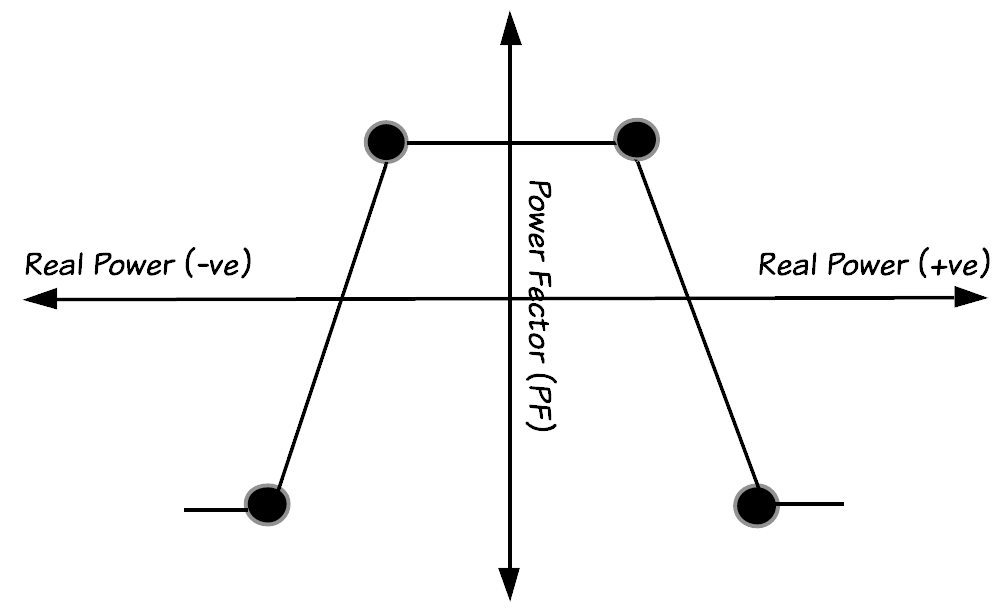}
    \caption{A typical curve Watt-Power Factor Function}.
    \label{fig:PFW}
\end{figure}

This inverter function adjusts the active power output of the inverter in response to the power factor value (which could range from 0 - 1, leading or lagging) of the system. The points defined in figure \ref{fig:PFW} are possible coordinates which defines the path for the watt-power factor control can be implemented.

According to \cite{EPRI2016}, some more future desirable SI functions include Energy storage direct charge/discharge management, Energy storage charge and discharge function based on the prevailing electricity price, coordinated charge and discharge of fleet of energy storage systems, load and generation following functions, dynamic event logging, coordinated status monitoring, ability to manage and monitor grid parameters with diverse configurations

\section{Proposed Methodology}
\begin{figure}[h!]
    \centering
\includegraphics[scale=0.21]{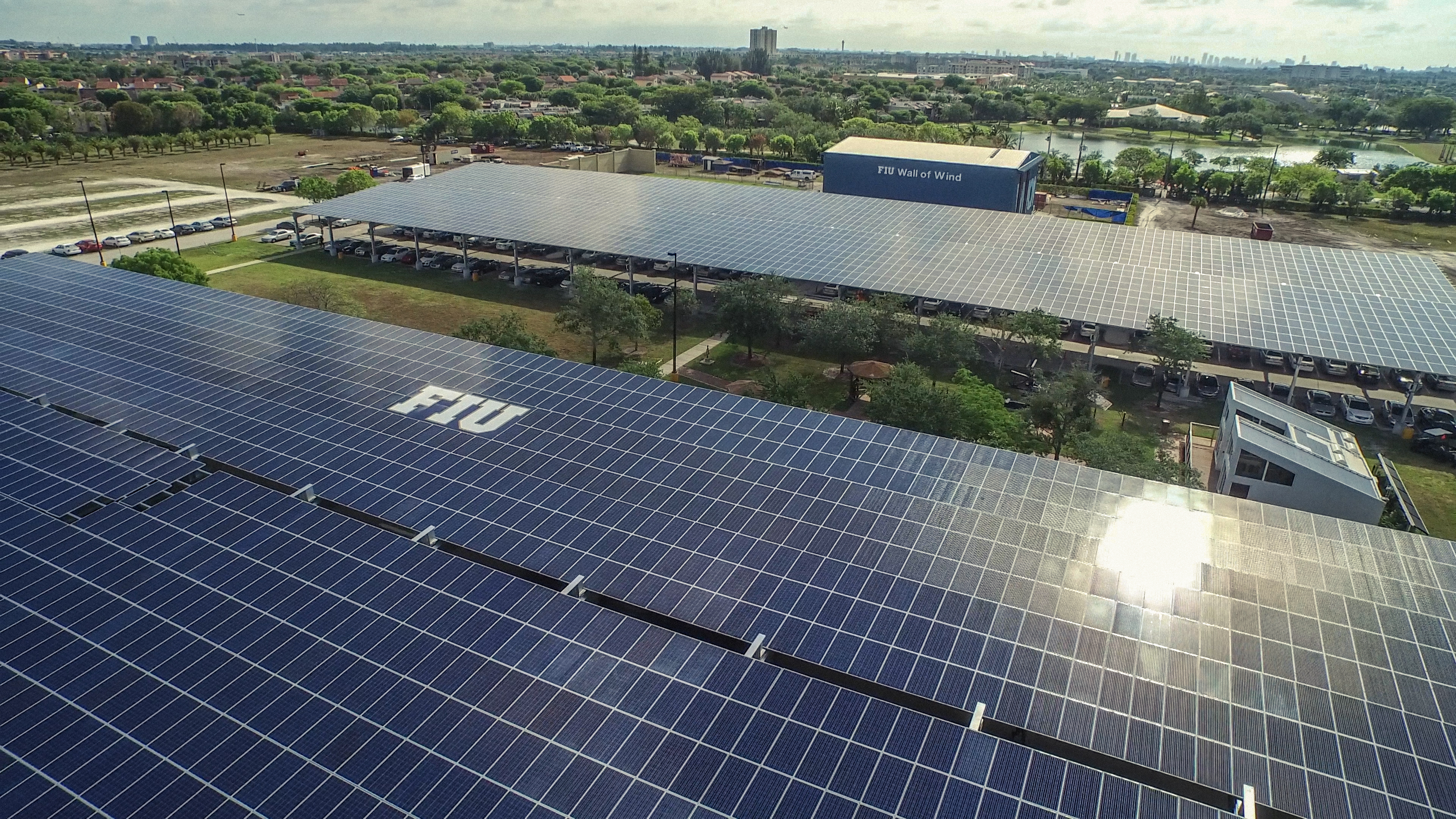}
    \caption{1.4 MW PV Located on FIU Engineering Campus used for data collection}
    \label{fig:PV_FIU}
\end{figure}

\begin{figure*}[t!]
    \centering
\includegraphics[scale=0.13 ]{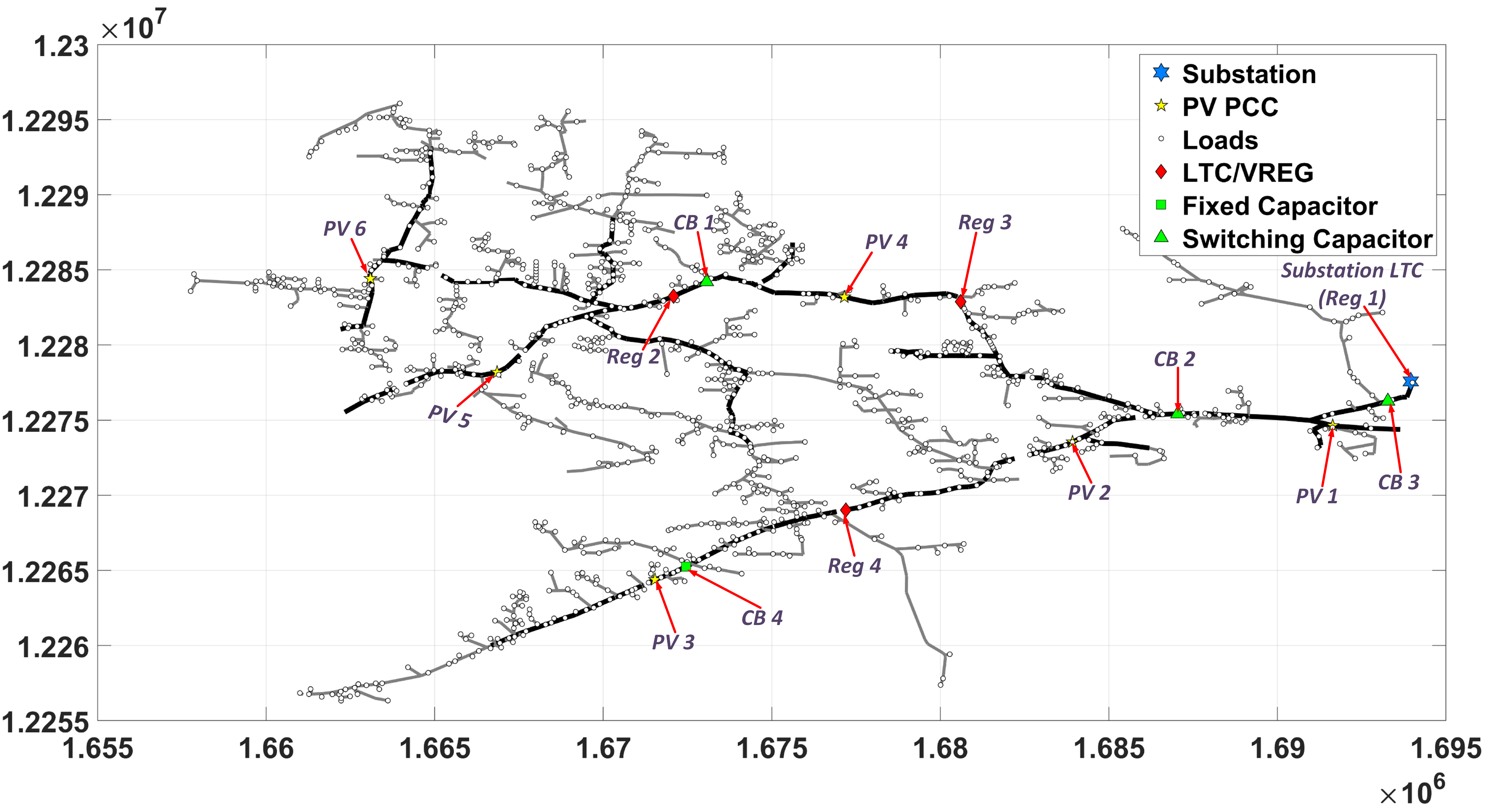}
    \caption{IEEE $8500$ Node test feeder with the integration of six PVs }.
    \label{fig:Feeder}
\end{figure*}

The SIs functionalities used in this study are those that significantly affect the voltage profile consequently the operations of the voltage control devices, the loses of the network as well as the system harmonics. Some of these functions are already being used by utility companies while others are new desirable functions that have been proposed. These are the volt-Watt, volt-Watt with rise/fall rate-of-change limit, volt-VAR, volt-VAR with adaptive set-point, volt-VAR with hysteresis, volt-VAR with low pass filter rate-of-change limit, maximum generation limit at 80 \%,  fixed power factor (FPF), dynamic reactive current injection/absorption. An IEEE $8500$ distribution test feeder developed by IEEE PES is used as a case study (as shown in Fig. \ref{fig:Feeder}). According to \cite{Arritt2010}, this feeder is a benchmark for distribution system analysis with elements that are typically found on North American feeders. The farthest node is from the feeder substation is approximately $17 km$. The feeder has four capacitor banks (three switched and one fixed), three VRs (Reg 2, Reg 3 and Reg 4) and one substation transformer LTC (Reg 1), balanced and unbalanced loads, feeder secondaries and service transformers. Six PVs are integrated into the network so as to be able to carefully observe the impacts of these SI settings on these legacy devices, the losses in the network, the harmonics and the CII. The modeling of the whole system is done using OpenDSS and MATLAB software.
Each of the PVs  has a rating of $1MW$ and SI rating of $1.2 MVA$. The irradiance profile used for this study was acquired from the actual irradiance sensor mounted on the $1.4 MW$ solar canopy installed in the Engineering Campus of Florida International University on a typical cloudy day as shown in Fig. \ref{fig:PV_FIU}. This is to enable adequate analysis of the impacts of the severe ramping on the various SIs settings used.
For the substation transformer OLTC (Reg 1) and the other three voltage regulators (Reg 2-4) in the system, the switching constraints is as expressed in (\ref{VRR}). Typically, these regulators have $32$ tap steps with $0.625\%$ difference in voltage per tap step \cite{adityaeclipse}. Commercially sold VRs are usually designed to make approximately $2$ million total number of switching for a typical life span of $20$ years. This constitutes a $273$ allowable tap steps per day \cite{Stiles2005,Manbachi2015}, which is expressed as a constraint in (\ref{TapLim}).

\begin{equation}
TS_{high} \geq TS_{t} \geq TS_{low}
\label{VRR}
\end{equation}

\begin{equation}
\sum _{t=1} ^ {T} SW_{taps,t} \leq SW_{max-tap,t}
\label{TapLim}
\end{equation}.

Equation (\ref{VRR}) ensures the tap step ($TS$) at time $t$ is within the highest and lowest allowable tap position of the VRs. Also, (\ref{TapLim}) limits the total number of tap switching steps with an interval $t$ to $T$ is within allowable maximum total number of switching steps within that period.

Similarly, the switching constraint for the capacitor banks is as expressed in (\ref{Lim2}) while (\ref{Qlim}). 

\begin{equation}
N_{DCS}=\sum _{t=1} ^ {T} N_{CS,t} \leq N_{CS}^ {max}
\label{Lim2}
\end{equation}.

\begin{equation}
 | Q_{cb max}^n| \geq |Q_{cb}^n|
\label{Qlim}
\end{equation}

Where $N_{DCS}$ is the total number of daily capacitor switching, $N_{CS,t}$ represents the number of capacitor switching at time t interval, $N_{{CS}^ {max}}$ is the daily allowable capacitor switching operations,  $ Q_{cb}^n$ is the amount of reactive power injected at node $n$ and $Q_{cb max}^n$ is the allowable reactive power injection at the same node $n$ \cite{Saleh,Tayo}.

The voltage constraint at each node is as expressed in (\ref{volconst}) according to ANSI
C84.1-2011 standard \cite{Tayo}.

\begin{equation}
\begin{split}
V_n^{max} \geq V_{n,t} \geq V_n^{min}\\
1.05 pu \geq V_{n,t} \geq 0.95 pu
\label{volconst}
\end{split}
\end{equation}
 where $V_n^{max}$, $V_{n,t}$, $V_n^{min}$ are the maximum ($1.05 pu$) node voltage, instantaneous node voltage and the minimum node voltage ($0.95 pu$) respectively. 
 
\subsection*{Economic Impact Analysis}
For simplicity, the cost of maintaining most voltage regulating switching devices in a power system network is usually assumed to be proportional to the number of times the  devices carries out its switching operations \cite{6466544,Cohen2015}. Different SIs setting affects the voltage profile of the distribution feeder differently, thereby causing the number of switching operations of these devices to differ.
A simplified way of quantifying the economic impact of the switching was presented by \cite{Thesis2013}. This is as expressed in (\ref{eq1}) and (\ref{eq2})

\begin{equation}
\label{eq1}
DCF= \frac{\text{Number of switching with PV}}{\text{Number of switching without PV}}
\end{equation}

\begin{equation}
\label{eq2}
OMC_{\text{with PV}}= OMC_{\text{without PV}} \times DCF
\end{equation}

\begin{equation}
\label{eq3}
CII= \frac{\sum OMC_{\text{with PV}} \text{(for all switching devices)}}{\sum  OMC_{\text{without PV}} \text{(for all switching devices)}}
\end{equation}

where DCF is the device cost factor, OMC is the operations and maintenance cost of the device and CII is the circuit impact index. A CII greater than unity shows an increase in the operation and maintenance cost of the system. This paper proposed the use of these economic indices to quantify the economic impacts of these various SIs settings on the network.
For this paper, the economic analysis will be done for the switching of the voltage regulators (Reg 2, Reg 3 and Reg 4) and the substation LTC (Reg 1). For the simplicity, operational and maintenance cost (OMC) of the VRs and LTC are assumed to be the same. And also, since the phases of the LTC and voltage regulator phases tap differently, the DCF will be computed for each VR and LTC as an aggregation of the number of tapping of each phases of the respective VRs and LTC.
For the feeder under study, the LTC (referred to as Reg 1) and three VRs (referred to as Reg 2, Reg 3 and Reg 4).

\begin{equation}
\label{eq4}
DCF_{Reg}= \frac {\sum Reg _{\text {phase A, B, C switching with PV}}}{\sum Reg _{\text {phase A, B, C switching without PV}}}
\end{equation}

\begin{equation}
\label{eq5}
OMC_{\textrm{Reg with PV}}= OMC_{\textrm{Reg without PV}} \times DCF_{Reg}
\end{equation}

Assuming $OMC_{\textrm{Reg without PV}}$ is the same for Reg 1, 2, 3 and 4; then the CII for the system can be simplified as:
\begin{equation}
\small
\label{eq6}
CII= \frac{\Big[DCF_{Reg 1}+DCF_{Reg 2}+DCF_{Reg 3}+ DCF_{Reg 4}\Big]_{\text {with PV}}} {\Big[DCF_{Reg 1}+DCF_{Reg 2}+DCF_{Reg 3}+ DCF_{Reg 4}\Big]_{\text {without PV}}}
\end{equation}

\section{Simulation Results and Analysis}
A time series simulation of the entire network with each SI function is done using OpenDSS coupled with Matlab. The irradiance used for the simulation is of 1-minute resolution with all the PV's SI set at the same function during each time series simulation.

A detailed summary of the various impacts of these SI settings on the switching and reactive power injection of the capacitor banks is as summarized in table \ref{table 2}.
Of all the nine SI functions considered, the Volt-Watt with rise/fall rate-of-change limiting  setting had the least impact on the average amount reactive power injection (CB 1=$430$ kVAR, CB 4=$335$ kVAR, no reactive power injection by CB 2 and CB 3).
The switching of the voltage regulator was highly impacted by the integration of these PVs. The irradiance profile used for the simulation has severe ramps in power generation from the PV which caused some frequent switching of the voltage regulators. Table \ref{table 4} shows the summary of number of switching carried out by each phase of the voltage regulators and substation LTC in the network.
The results showed that the FPF setting tapped the most while Volt-Watt with rise/fall rate-of-change limit tapped the least. The FPF function forces the substation transformer to tap many times which rarely occurred with other settings. The high voltage regulator operations of FPF setting can also attributed to the fact that the SI did little voltage control with most of the voltage regulation and control carried put by the LTC, VRs and the capacitor banks.

\begin{table*}[h!]
\caption{Summary of Capacitor Banks reactive power injection.}
\centering
%\small
\label{table 2}
\begin{center}
\begin{tabular}{|c|c|}
\hline
 \centering{\textbf{ Setting}}&{\textbf{Impacts on CB 1, CB 2, CB 3, and CB 4}} \\
 \hline
 Volt-Watt &
 \multicolumn{1}{m{10cm}|}{There was approximately no reactive power injection from CB 2 and CB 3. This is possibly due to their proximity to the substation and the stiffness of the voltage profile close to the substation. CB 1 and CB 4 downstream injected large amount of reactive power with an average of 440 kVAR and 330 kVAR per phase respectively.}\\

\hline
 Volt-Watt with rise/fall rate-of-change limit &\multicolumn{1}{m{10cm}|}{No reactive power injection by CB 2 and CB 3. CB 1 and CB 4 injected an average of 430 kVAR and 335 kVAR respectively. The limit to the rate of change of active power generation by the PV, limited the fluctuation in the reactive power injection when there is severe ramp up and down by the PV.}  \\
\hline
  Volt-VAR&\multicolumn{1}{m{10cm}|}{ No reactive power injection by CB 2 and CB 3. The rate of switching of CB 1 and CB 4 was more than that of the Volt-Watt setting. This is due to the significant impact of this setting on the feeder voltage profile from the PVs. There is a slight increase in reactive power injection using this setting compared to the Volt-Watt}	\\
 \hline
 Volt-VAR with adaptive set-point& \multicolumn{1}{m{10cm}|}{In comparison with the ordinary Volt-VAR, the adaptive set-point caused  CB 1 to inject zero reactive power in phases B and C and some reactive power during a few minutes in phase A. Reactive power injection by CB 4 is similar to that of the Volt-VAR. The adaptive set-point led to few more taps with higher tap value compared to the Volt-VAR}\\
 \hline
  Volt-VAR with hysteresis & \multicolumn{1}{m{10cm}|}{The adaptive set-point settings lead to  fluctuation in reactive power injection when there was no generation from the PVs. The amount of reactive power injection from CB 1 and CB 4 was similar to the of the Volt-VAR. Phase C of CB 1  and CB 3 showed some exception in the kVAR injection. Phase C of CB1 had a relatively constant kVAR with a sudden reduction to zero at some points and phase C of CB 3  had zero injection and approximately 330 kVAR injection at some points during the simulation.}\\
\hline

 Volt-VAR with low pass filter rate-of-change limit & \multicolumn{1}{m{10cm}|}{Lesser rate of reactive power injection by CB 1 and CB 4 due to the low pass filter settigs. Phase C of CB 1 was zero for most of the period with sudden increase to 440 kVAR for some minutes during the time interval.} \\
\hline
  Maximum generation limit at 80 \%& \multicolumn{1}{m{10cm}|}{Voltage control was solely done by the capacitor banks and voltage regulators. The reactive power injection by CBs 1,2,3 and 4 is similar to that of the Volt-Watt with a more flattened (less fluctuating) reactive power injection. }\\
\hline
FPF setting at 0.8 & \multicolumn{1}{m{10cm}|}{PVs appeared slightly appears capacitive with fixed reactive power injection. The kVAR injection by CB 4 is more compared to Volt-Watt and Volt-VAR. This is partly due to the absence of voltage control by the PVs. CB 1 has a sharp decrease of its kVAR to zero during period of power generation from the PVs. Also, phase B of CB 3 injected some kVAR at some instance during the time interval.}\\
 \hline
  Dynamic reactive current injection/absorption& \multicolumn{1}{m{10cm}|}{Similar to the Volt-VAR but with some increase in slope in VAR output on phase B of CB 3 at some time during PV generation. Also on the average, the reactive power injection by CB 1 and CB 4 is slightly higher compared to the Volt-VAR setting.} \\
\hline
\end{tabular}
\end{center}
\end{table*}

\begin{table*}[h!]
\caption{Number of tap changes with each setting}
\centering
%\small

\label{table 4}
%\begin{tabular}{|c|c|c|c|c|c|c|c|c|c|c|c|c|}
\begin{tabular}{|p{0.33\linewidth}| p{0.027\linewidth}|p{0.027\linewidth}|p{0.027\linewidth}|p{0.027\linewidth}|p{0.027\linewidth}|p{0.027\linewidth}|p{0.027\linewidth}|p{0.027\linewidth}|p{0.027\linewidth}|p{0.027\linewidth}|p{0.027\linewidth}|p{0.027\linewidth}|}

\hline

\centering{\multirow{4}{*}{\textbf{Settings}}}&\multicolumn{12}{c|}{\textbf{VR (Phases)}}\\
\cline{2-13}
& \multicolumn{3}{c|}{VR 1}&\multicolumn{3}{c|}{VR 2}&\multicolumn{3}{c|}{VR 3}&\multicolumn{3}{c|}{VR 4}\\
\cline{2-13}
 & A&B&C&A&B&C &A &B&C & A&B&C\\
\hline
  \centering{ Volt-Watt}&0 & 0&0 & 131& 97& 23& 126& 122 & 91 & 145& 137&65 \\
  \hline
  \centering{Volt-Watt with rise/fall rate-of-change limit}&0 &0&0&0&1& 0&1 &1& 1&2 &2&2\\ 
  \hline
  \centering{Volt-VAR}&0 &0&0&125&119& 83& 148&151&59 &146 &158&98\\ 
  \hline
  \centering{ Volt-VAR with adaptive set-point}& 0&0&0&106&102&39 &105 &85& 30&128 &146&76\\
  \hline
  \centering{Volt-VAR with hysteresis}&0 &16&21&114&84& 77&148 &142&97 &160 &161&123\\ 
  \hline
  \centering{Volt-VAR with low pass filter rate-of-change limit}&0 &0&8&147&112& 62& 135&137& 54&158 &144&124\\ 
  \hline
  \centering{ Maximum generation limit at 80 \%}&0 &0&0&150&109&61 & 144&141& 92&162 &170&111\\ 
  \hline
  \centering{FPF setting at 0.8}& 0&126&126&193&186& 162& 234&231& 228&234 &228&223\\ 
  \hline
  \centering{Dynamic reactive current injection/absorption}&0 &4&0&150&105&104 &145 &137&89 & 178&155&114\\ 

\hline
\end{tabular}

\end{table*}

\begin{figure*}[h!]
    \centering
\includegraphics[scale=0.17]{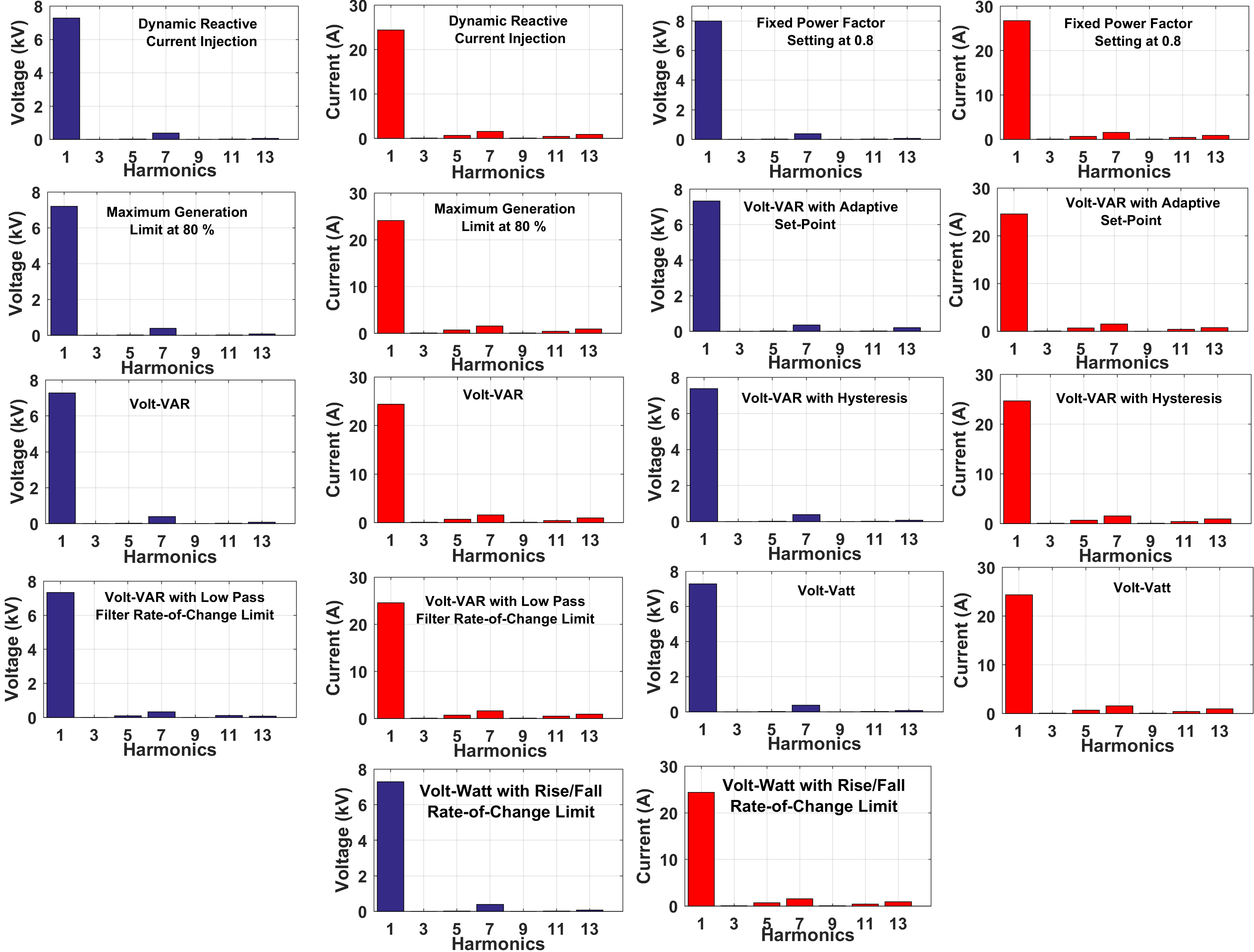}
    \caption{The harmonic measurement from a line with different SI settings}.
    \label{Harmonic}
\end{figure*}

The harmonic content of the voltage and current flowing through one of the lines in the network was monitored to investigate the impact of these settings on the voltage and current harmonics of the system at this point of reference. Figure \ref{Harmonic} shows the $1^{st}$, $3^{rd}$, $5^{th}$, $7^{th}$, $11^{th}$ and $13^{th}$ harmonic components of the voltage and current measurement. 
The plots shows that the current harmonic of the system was rarely impacted by the different settings used. The current harmonics with the different SI settings is almost the same for the feeder under study. However, the voltage harmonics showed some variations in the harmonic component. The Volt-VAR with adaptive set-point had a slightly higher $13^{th}$ harmonic components compared to the  others setting. The Vol-VAR with the low pass filter rate-of-change limit settings had slightly more $5^{th}$ harmonic component compared to the other settings.

\begin{figure}[h!]
   \centering
\includegraphics[scale=0.12]{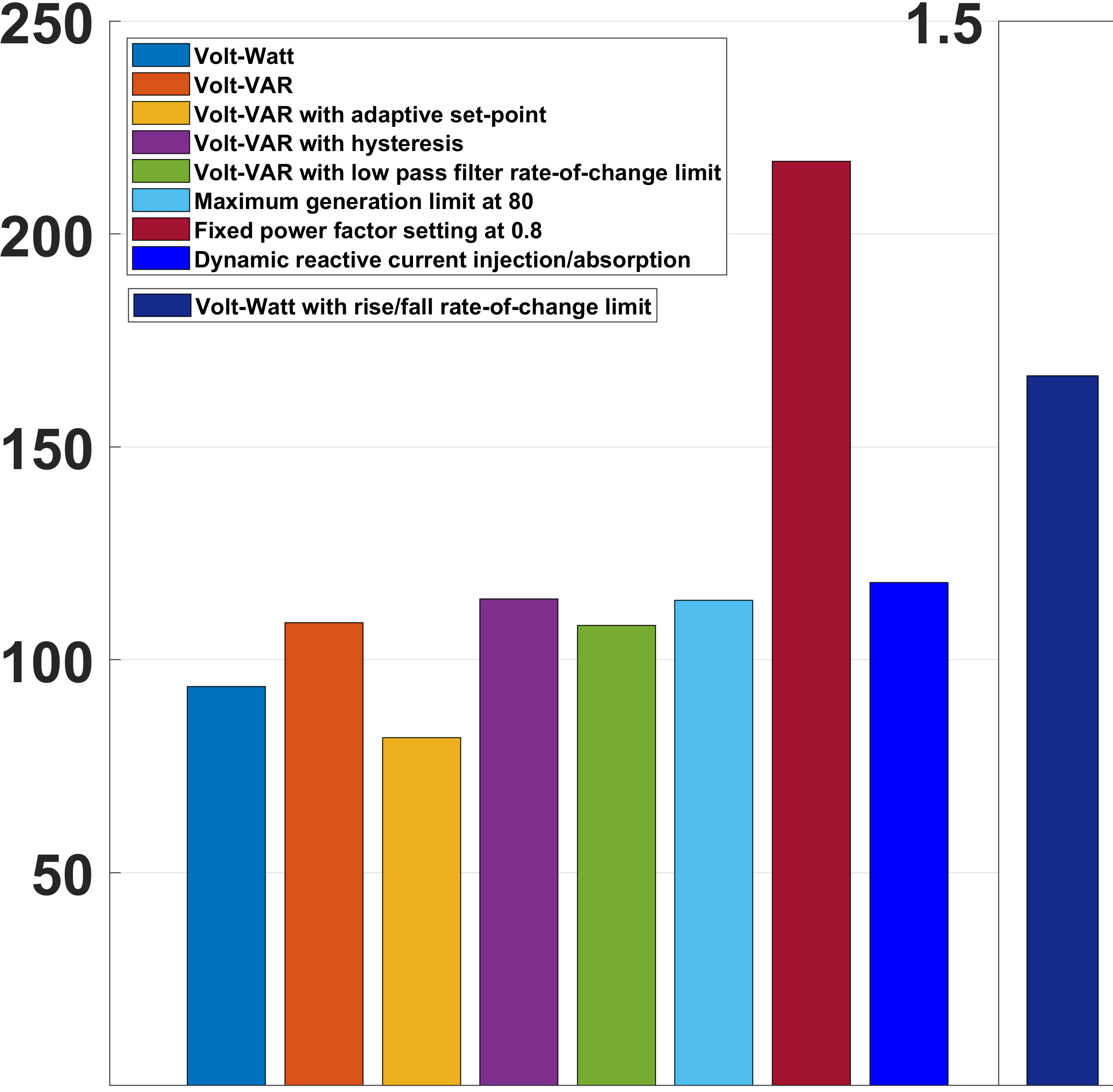}
    \caption{The system CII with different SI functionalities}.
   \label{fig:CII}
\end{figure}
\begin{figure}[h!]
   \centering
\includegraphics[scale=0.06]{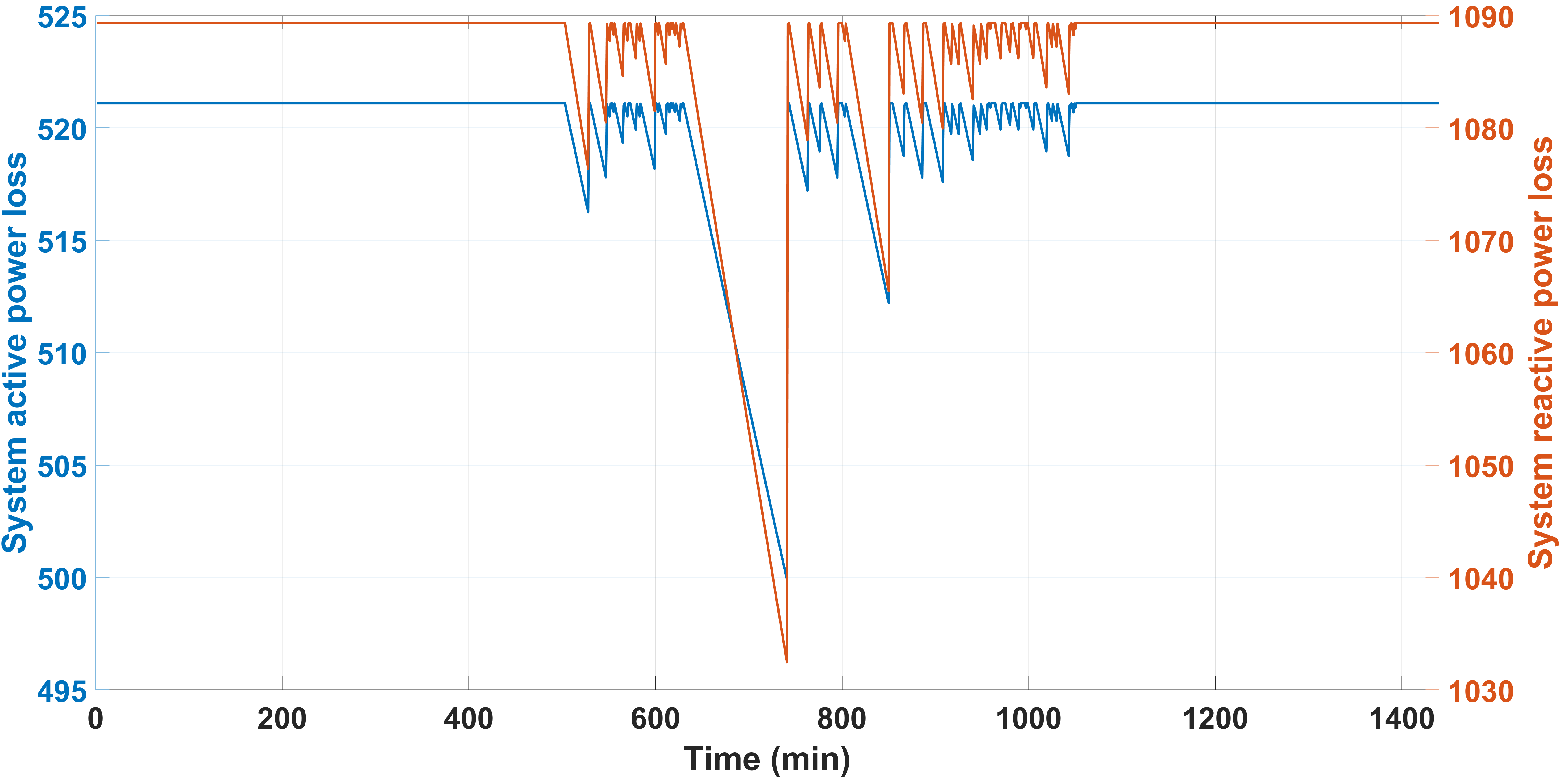}
   \caption{Volt-Watt with rise/fall rate-of-change limiting loss profile }
   \label{fig:LossVWRFCL}
     
\end{figure}

Figure \ref{fig:CII} shows the plot of the CII for the various SI setting used in the simulation. As previously mentioned, a high CII values indicates the an increase in operations and maintenance cost of the voltage regulators. The Volt-Watt with rise/fall rate-of-change limiting  setting showed the least impact in terms of the CII ($1.5$).
The FPF setting ($0.8 pu$) has the highest value of CII. This can be explained by the fact that the PV appears capacitive to the network with constant reactive power injection irrespective of the voltage profile of the network which leads to higher number of switching of the voltage regulators and the LTC. The least value of CII is seen with the Volt-Watt rise/fall rate of change limit setting. This functionality constantly limits the ramping of the PV power which usually necessitates more switching by the voltage regulators.
For all the settings considered, the system active and reactive power losses were reduced during the PV generation. The least amount of reduction in system losses (both active and reactive) was observed with the Volt-Watt with rise/fall rate-of-change limiting setting as shown in ($Q_{loss}=1090$ kVAR and $P_{loss}=500$ kW) Fig. \ref{fig:LossVWRFCL}.

\begin{figure}[h!]
   \centering
  \includegraphics[scale=0.06]{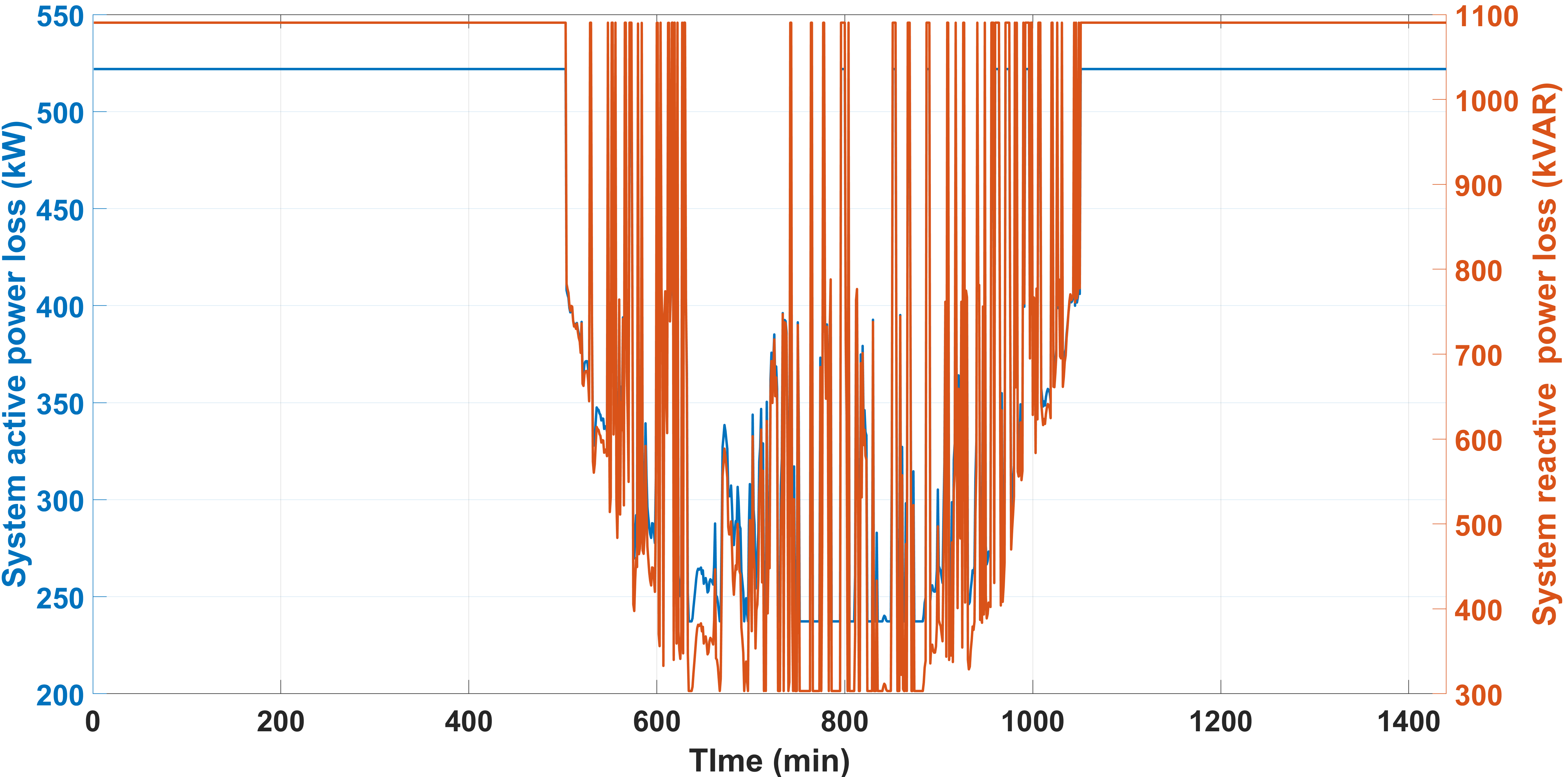}
     \caption{Maximum generation limit at 80 \% loss profile}
  \label{Maxgen}
    
\end{figure}

This explains why there are few taps in the voltage regulator and few switching in capacitor banks while this setting was used. The setting had a minimal impact on the voltage profile of the network during the simulation. The Maximum generation limit setting caused the highest reduction in system active power losses as shown in Fig. \ref{Maxgen}.

\section{Conclusion}
This paper presented a comprehensive analysis of some SI functionalities using the standard IEEE $8500$ node test feeder modeled. The results showed the various impacts of these SIs functions on the operation of voltage regulators, capacitor banks, the losses in the system, the circuit impact index and their possible impact on the harmonics components of the network. Different functions of the SIs impacted on the feeder in different ways. The Volt-Watt with rise/fall rate-of-change limiting  setting showed the least impact in terms of the CII, total voltage regulator tapping and reactive power injection by capacitors, but with highest amount of losses compared to other SI functions.
Though every system is unique depending on the network structure and the existing infrastructure, and impacts could be feeder specific, this paper provides an insight on how the evaluated SI functions impact the standard and most complex IEEE $8500$ node distribution feeder considered. Formulation of voltage optimization algorithms  and optimal settings of these smart inverter curves will be required is necessary for effective voltage control.

\ifCLASSOPTIONcaptionsoff
  \newpage
\fi

%\bibliographystyle{IEEEtran}

\end{document}